\documentclass[12pt]{iopart}
\usepackage{iopams}

\newcommand{\CD}{{\cal D}}
\newcommand{\CR}{{\cal R}}
\newcommand{\average}[1]{\left\langle #1 \right\rangle_\CD}

\newcommand{\gaverage}[1]{\left\langle #1 \right\rangle_{\Sigma}}
\newcommand{\initial}[1]{{#1_{\rm \bf i}}}
\newcommand{\inI}{{I}}
\newcommand{\inII}{{II}}

\begin{document}

\title[A cosmic equation of state for the inhomogeneous Universe]{A cosmic equation of state 
for the inhomogeneous Universe: can a global far--from--equilibrium state explain Dark Energy?}
\author{Thomas Buchert}

\address{Arnold Sommerfeld Center for Theoretical Physics,
Ludwig--Maximilians--Universit\"{a}t, Theresienstra{\ss}e 37,
80333 M\"{u}nchen, Germany \\Email: buchert@theorie.physik.uni-muenchen.de}

\begin{abstract}
A system of effective Einstein equations for spatially averaged scalar variables 
of inhomogeneous cosmological models can be solved by providing a 
`cosmic equation of state'. 
Recent efforts to explain Dark Energy focus on `backreaction effects' of inhomogeneities 
on the effective evolution of cosmological parameters in our Hubble volume, avoiding a
cosmological constant in the equation of state.  
In this {\it Letter} it is argued that, if kinematical 
backreaction effects are indeed of the order of the
averaged density (or larger as needed for an accelerating domain of the Universe),
then the state of our regional Hubble volume would have to be in the vicinity of a 
far--from--equilibrium state that balances kinematical backreaction and average density. 
This property, if interpreted globally, is shared by a stationary cosmos with 
effective equation of state ${\bf p}_{\bf\rm eff} = -1/3 \,\boldsymbol{\varrho}_{\bf\rm eff}$.
It is concluded that a confirmed explanation of Dark Energy by kinematical backreaction 
may imply a paradigmatic change of cosmology.
\end{abstract}



\section{Effective Einstein equations and the cosmic equation of state}

To set notation and to provide the framework for our argument, 
we recall a set of effective Einstein equations \cite{buchert:grgdust}. 
The argument presented can be carried over to studies of 
inhomogeneous cosmologies covering the Early Universe and radiation--dominated epochs
with the help of the more general effective equations developed in \cite{buchert:grgfluid}.

\subsection{Averaged equations}

For the sake of transparency we restrict ourselves to the matter model {\it irrotational dust}.
Adopting a foliation of  spacetime into flow--orthogonal hypersurfaces (which is possible for
irrotational dust) with the 3--metric $g_{ij}$ in the line--element 
$ds^2 = -dt^2 + g_{ij}\,dX^i dX^j$, we define spatial averaging of a scalar field $\Psi$ on 
a domain $\CD$ with volume $V_\CD$ by: 
\begin{equation}
\label{eq:average-GR}
\average{\Psi (t, X^i)}: = 
\frac{1}{V_\CD}\int_\CD J d^3 X \;\;\Psi (t, X^i) \;\;,
\end{equation}
with  $J:=\sqrt{\det(g_{ij})}$, where  $g_{ij}$ is  the metric  of the
spatial  hypersurfaces, and  $X^i$ are  coordinates that  are constant
along  flow lines. 
Following \cite{buchert:grgdust} we define an 
{\it effective scale factor} through the volume of a simply--connected domain 
$\CD$ in a hypersurface, normalized by the  volume of the initial domain $\initial\CD$,
$a_\CD := \left(V_\CD / V_{\initial\CD}\right)^{1/3}$.
Employing the fact that, for a restmass preserving domain $\CD$, 
volume averaging of a scalar function $\Psi$
does not commute with its time--evolution, $\;\;\average{\partial_t \Psi} - \partial_t 
{\average{\Psi}}\;=\;\average{\Psi}\average{\theta} -\average{\Psi\theta}\;\;$,
we can derive an effective equation for the spatially averaged expansion 
$\average{\theta} \;=\;\frac{{\dot V}_\CD}{V_\CD}\;=\;3\frac{{\dot a}_\CD }{a_\CD }
\;=\;: 3 H_\CD$ (with an {\it effective Hubble functional} $H_\CD$
(an overdot denotes partial time--derivative): 
setting $\Psi \equiv \theta$, inserting {\it Raychaudhuri's evolution equation} 
for $\partial_t \theta$ into the commutation rule above,
and using the effective scale--factor $a_\CD$ one obtains:
\begin{equation}
\label{averageraychaudhuri}
3\frac{{\ddot a}_\CD}{a_\CD} + 4\pi G 
\average{\varrho} - \Lambda = {\cal Q}_\CD\;\;\;;\;\;\;
\average{\varrho}=\frac{M_\CD}{V_{\initial\CD}a_\CD^3}\;\;.
\end{equation}
The first integral of (\ref{averageraychaudhuri}) is directly given by averaging 
the {\it Hamiltonian constraint}:
\begin{equation}
\label{averagehamilton}
3\left( \frac{{\dot a}_\CD}{a_\CD}\right)^2 - 8\pi G\average{\varrho}
+ \frac{\average{\CR}}{2} 
- \Lambda = -\frac{{\cal Q}_\CD}{2} \;\;,
\end{equation}
where   the  total restmass   $M_\CD$,  the averaged density, the  averaged  spatial  Ricci  scalar
$\average{\CR}$   and   the  {\it kinematical backreaction  term}  ${\cal Q}_\CD$   are
domain--depen\-dent and, except the mass, time--depen\-dent functions.
The backreaction source term is given by
\begin{equation}
\label{eq:Q-GR} 
{\cal Q}_\CD : = 2 \average{\inII} - \frac{2}{3}\average{\inI}^2 \;=\;
\frac{2}{3}\average{\left(\theta - \average{\theta}\right)^2 } - 
2\average{\sigma^2}\;\; ;
\end{equation}
here,  $\inI = \Theta^i_{\;i}$  and $\inII = \frac{1}{2}[\,(\Theta^i_{\;i})^2 - 
\Theta^i_{\;j}\Theta^j_{\;i}\,]$  
denote  the  principal scalar invariants  of the  expansion
tensor, defined  as minus the extrinsic
curvature  tensor, $-K_{ij}:=\Theta_{ij}$. In the second equality above it was split 
into kinematical invariants through the decomposition  
$\Theta_{ij} = \frac{1}{3}g_{ij}\theta + \sigma_{ij}$, with the rate of expansion 
$\theta =\Theta^i_{\;i}$ and
the rate of shear $\sigma^2 = \frac{1}{2}\sigma_{ij}\sigma^{ij}$. (Note that vorticity is absent in
the present gauge; we adopt the summation convention.)

The time--derivative of the averaged Hamiltonian constraint (\ref{averagehamilton})
agrees with the averaged Raychaudhuri equation (\ref{averageraychaudhuri}) by virtue of the
following {\it integrability  condition}:
\begin{equation}
\label{integrability}
\fl
\partial_t {\cal Q}_\CD + 6 H_\CD {\cal Q}_\CD +  
\partial_t \average{\CR} + 2 H_\CD \average{\CR} = 0 \;\Leftrightarrow \;
\partial_t \left(\,{\cal Q}_\CD \, a_\CD^6 \,\right) + a_\CD^{4}\;
\partial_t \left(\,\average{\CR}a_\CD^2 \,\right)=0\;.
\end{equation}

\subsection{The cosmic quartet}

We may further introduce dimensionless average characteristics as follows 
\cite{buchert:grgdust}:
\begin{equation}
\label{omega}
\fl\qquad
\Omega_m^{\CD} : = \frac{8\pi G \average{\varrho}}{3 H_{\CD}^2 } \;\;;\;\;
\Omega_{\Lambda}^{\CD} := \frac{\Lambda}{3 H_{\CD}^2 }\;\;;\;\;
\Omega_{R}^{\CD} := - \frac{\average{\cal R}}{6 H_{\CD}^2 }\;\;;\;\;
\Omega_{Q}^{\CD} := - \frac{{\cal Q}_{\CD}}{6 H_{\CD}^2 } \;\;,
\end{equation}
where we have employed the  effective Hubble--functional $H_\CD$ introduced above
that reduces to Hubble's function in the homogeneous--isotropic case.
With these definitions the averaged Hamiltonian constraint (\ref{averagehamilton}) reads:
\begin{equation}
\label{hamiltonomega}
\Omega_m^{\CD}\;+\;\Omega_{\Lambda}^{\CD}\;+\;\Omega_{R}^{\CD}\;+\;
\Omega_{Q}^{\CD}\;=\;1\;\;,
\end{equation}
providing a scale--dependent {\it cosmic quartet} relating all relevant 
``cosmological parameters''. For ${\cal Q}_\CD = 0$ the above functionals 
reduce to the corresponding parameters of the standard homogeneous--isotropic
models.

The effective cosmological parameters defined in
(\ref{omega}) can be considered to provide a fair representation of the values which also an 
observer would measure in a sufficiently shallow survey region $\CD$ (the light--cone effect is not taken
into account). We may therefore discuss estimates of those parameters in comparison with
observed values.  Note, however, that the {\it interpretation} of observations is mostly done
by employing a standard Friedmannian cosmology as a `fitting model' and therefore, 
geometrical inhomogeneities (that are hidden in the definition of the spatial averages 
in the Riemannian volume element, {\it cf.} Eq.~(\ref{eq:average-GR}))
are ignored \cite{buchertcarfora:PRL}.

\subsection{The cosmic equation of state}

The above equations can formally be recast into standard zero--curvature Friedmann equations 
with new effective sources (\cite{buchert:grgfluid}:~{\sl Corollary 2})\footnote{Note that in this representation
of the effective equations ${\bf p}_{\bf\rm eff}$ just denotes
a formal ``pressure'':  in the perfect fluid case with an inhomogeneous pressure function
the foliation has to be differently chosen and there is a further averaged pressure gradient
term \cite{buchert:grgfluid}.}:
\begin{equation}
\label{equationofstate}
\fl
\boldsymbol{\varrho}^{\CD}_{\bf\rm eff} = \average{\varrho}-\frac{1}{16\pi G}{\cal Q}_\CD - 
\frac{1}{16\pi G}\average{\CR}
\;\;\;;\;\;\;{\bf p}^{\CD}_{\bf\rm eff} =  -\frac{1}{16\pi G}{\cal Q}_\CD + \frac{1}{48\pi G}\average{\CR}\;\;.
\end{equation}
\begin{equation}
\label{effectivefriedmann}
\fl
3\frac{{\ddot a}_\CD}{a_\CD} =
\Lambda - 4\pi G (\boldsymbol{\varrho}^{\CD}_{\bf\rm eff}
+3{\bf p}^{\CD}_{\bf\rm eff})\;;\;
3H_\CD^2 =\Lambda + 8\pi G 
\boldsymbol{\varrho}^{\CD}_{\bf\rm eff}\;;\;
{\dot{\boldsymbol{\varrho}}}^{\CD}_{\bf\rm eff} + 
3H_\CD \left(\boldsymbol{\varrho}^{\CD}_{\bf\rm eff}
+{\bf p}^{\CD}_{\bf\rm eff} \right)=0\;.
\end{equation}
Eqs.~(\ref{effectivefriedmann}) correspond to the equations (\ref{averageraychaudhuri}),
(\ref{averagehamilton}) and (\ref{integrability}), respectively.
In these equations the kinematical backreaction term ${\cal Q}_\CD$ itself 
obeys a {\it stiff} equation of state mimicking a dilatonic field in the fluid analogy
(for further implications see \cite{buchert:grgfluid}).

Given an equation of state in the form ${\bf p}^\CD_{\bf\rm eff} = 
\beta (\boldsymbol{\varrho}^\CD_{\bf\rm eff}, a_{\cal D})$ that relates the effective 
sources (\ref{equationofstate}), 
the effective Friedmann equations (\ref{effectivefriedmann})
can be solved (one of the equations (\ref{effectivefriedmann}) is then redundant).  
Therefore, any question posed that is related to the evolution of scalar characteristics
of inhomogeneous universe models may be ``reduced'' to finding the {\it cosmic state}
on a given spatial scale. Although
formally  similar to the situation in Friedmannian cosmology, here the equation of state
is {\it dynamical} and depends on details of the evolution of inhomogeneities.
In general it describes non--equilibrium states.

\section{Explaining Dark Energy through kinematical backreaction}

The `coincidence' that a Dark Energy source (modeled in the simplest case by a cosmological
constant) starts to dominate around the epoch when also
structure enters the non--linear regime suggests that there could be
a physical  relation between the effect of structure on the average
expansion (known as {\it backreaction effect}) and the Dark Energy gap found in the
Friedmannian standard model, thus providing a natural solution to this {\it coincidence problem}
(besides the fact that more ``exotic'' explanations are then not needed). 

The averaging problem
in cosmology has a long history including calculations of the backreaction
effect  shortly after George Ellis \cite{ellis} has pointed out its importance (\cite{futamase1}
and many works thereafter; references may be found in 
\cite{bks,kolbetal,rasanen,ellisbuchert,buchert:static}).
The new input into this discussion is the (not unsupported) 
claim by Edward Kolb et al. \cite{kolbetal} 
that the condition 
(\ref{accelerationcondition}) below could be satisfied 
within our regional Hubble volume, hence providing a smart explanation of
the Dark Energy problem \cite{wetterich} without employing a cosmological constant,
quintessence or corrections to Einstein's laws of gravity.
 
Setting $\Lambda = 0$ in Eq.~(\ref{averageraychaudhuri}), 
the condition for an accelerating patch $\CD$ of the Universe directly follows:
\begin{equation}
\label{accelerationcondition}
{\cal Q}_\CD \;>\;4\pi G \average{\varrho}\;\;\;;\;\;\; \average{\varrho}= 
\frac{M_\CD}{V_{\initial\CD}a_\CD^3}  \;\ge 0      \;\;.
\end{equation}
(With regard to (\ref{eq:Q-GR}), in order for ${\cal Q}_\CD$ to be {\it positive}, 
expansion fluctuations would have to dominate over shear fluctuations.)
This regional condition is weaker than the requirement of global acceleration, since it 
accounts for the regional nature of our observations. 
There is, however, a large body of opponents including myself who do not think that
the condition (\ref{accelerationcondition}) can be met within the standard picture of 
structure formation from CDM initial conditions.
A number of caveats would have to be overcome
related to explicit calculations of kinematical backreaction and observational constraints,
which both will be discussed in more detail in a forthcoming paper \cite{buchert:static}. 
However, those caveats largely
depend on  assumptions that would consider perturbation theory on a
Friedmannian background, would extrapolate ({\ref{accelerationcondition}) to the global
scale according to the cosmological principle, and also would ignore the difficulties in relating the
model parameters (\ref{omega}) to observations \cite{ellisbuchert}. 
It is important to keep this disclaimer in mind in what follows. 
Restricting our attention to the universe {\it model}, as we shall do,
there is no such caveat in our line of arguments.

Let us give an example that illustrates 
how strong the condition (\ref{accelerationcondition}) appears,
if we straightly compare the model parameters (\ref{omega}) with current 
observations. We first rewrite the condition (\ref{accelerationcondition})
in terms of the dimensionless characteristics (\ref{omega})\footnote{Note that, for a 
positive ${\cal Q}_{\CD}$, $\Omega_{Q}^{\CD}$ is, by definition, negative.}:
\begin{equation}
\label{accelerationconditionomega}
-\Omega_{Q}^{\CD} \;>\;\frac{\Omega_m^{\CD}}{4}\;\;.
\end{equation}
We have to be aware that, if (\ref{accelerationconditionomega}) holds on some large
domain $\CD$, which we may take to be 
as large as our observable Universe \cite{kolbetal}, then Hamilton's constraint
in the form (\ref{hamiltonomega}) also implies: 
\begin{equation}
\label{curvaturecondition}
\Omega_{\Lambda}^{\CD} \;+\;\Omega_{R}^{\CD}\;>\;1\;-\;\frac{3}{4}\,\Omega_m^{\CD}\;\;,
\end{equation}
showing that, for a low density parameter, we would need a substantial amount of 
negative curvature (positive $\Omega_{R}^{\CD}$)
in the inhomogeneous model (not in the `fitting model') 
on the domain $\CD$, if we put the 
cosmological constant equal to zero. 
(To reconcile a small curvature parameter in this condition with a non--vanishing
cosmological constant would need
an even larger value of $\Omega_{\Lambda}^{\CD}$ than that suggested by the 
`concordance model' of about $0.7$.)
The fact that a large value of kinematical backreaction goes along with 
a substantial amount of average Ricci
curvature has also been stressed and discussed by Syksy R\"as\"anen \cite{rasanen}.
The condition (\ref{curvaturecondition}), taken at face value (there are arguments why this
could be naive),  would contradict the widely agreed expectation that the
curvature should be very small 
(the `concordance model' assumes an exactly zero scalar curvature).

\smallskip

We, here, do not attempt to respectively verify or falsify the condition 
(\ref{accelerationcondition}), but instead
follow a line of arguments that assumes (\ref{accelerationcondition}) to hold.
That is to say, {\it if} kinematical backreaction can indeed account for
the Dark Energy gap found in Friedmannian cosmology, then we have to closely 
examine implications of the above condition. 

Let us first remark that the physical contents of  (\ref{accelerationcondition}) implies 
a strongly fluctuating cosmos, roughly speaking: fluctuations encoded in ${\cal Q}_\CD$ have 
to be of the same order as the average density. 
This can only
happen, if there is a strong coupling of kinematical backreaction to the averaged scalar 
curvature, even on the Hubble scale, for if this coupling is absent, 
(\ref{integrability}) admits the special solution 
$\average{\CR} \propto a_\CD^{-2}$ and ${\cal Q}_\CD \propto a_\CD^{-6}$ 
(\cite{buchert:grgdust} App.B),
i.e. averaged scalar curvature behaves as in a constant--curvature Friedmannian model 
with effective scale factor $a_\CD$, and
fluctuations decay with the square of the (expanding) volume capturing what 
we may call {\it cosmic variance}. This particular solution mirrors what
we would expect from standard cosmology. It implies that the averaged density 
$4\pi G \average{\varrho} \propto a_\CD^{-3}$ would substantially dominate over 
${\cal Q}_\CD  \propto a_\CD^{-6}$
in a globally expanding universe model after some time, even if we would start with the
condition (\ref{accelerationcondition}). 
Therefore, a strong coupling 
of  ${\cal Q}_\CD$ to $\average{\CR}$ that changes the dependence on the effective
scale factor sufficiently, is key to the explanation of Dark Energy through 
kinematical backreaction.

Notwithstanding, if we assume such a strong coupling exists in a realistic universe model, 
and if we suppose that our Hubble volume accelerates due to the fact that 
${\cal Q}_\CD$ dominates over $4\pi G \average{\varrho}$, 
then we are entitled to say that a {\it typical} Hubble volume
would correspond to a non--perturbative state in the vicinity
of ${\cal Q}_\CD \approx 4\pi G \average{\varrho}$, i.e. it would {\it not} correspond to a 
perturbative state
in the vicinity of a model with ${\cal Q}_\CD \approx 0$, as expected in the 
standard picture of small perturbations of a Friedmannian background. 

We are now going to identify this state in the effective equations.
For this end let us now extend the spatial domain $\CD$ to the whole 
Riemannian manifold $\Sigma$, which we assume to be compact.
The cosmological principle would extrapolate the condition (\ref{accelerationcondition})
to the global scale. 
However, since we are assuming a strongly fluctuating cosmos, it is more appropriate  
to allow for other Hubble volumes
that are slightly decelarating with ${\cal Q}_\CD \;<\; 4\pi G \average{\varrho}$, so that a 
{\it typical} Hubble volume 
would reside in a state close to the balance condition
${\cal Q}_\CD \;=\; 4\pi G \average{\varrho}$\footnote{In \cite{buchert:static} a
conservative estimate, based on current observational parameters, 
shows that such a cosmos provides room for at least $50$ Hubble volumes.}. 
This balance condition furnishes an example for such a state.
Extrapolating this condition to the
global scale implies with (\ref{averageraychaudhuri}) that 
the global effective acceleration vanishes, and 
together with  (\ref{averagehamilton}) we face the {\it global stationarity conditions}:
\begin{equation}
\label{stationary}
\fl\quad
{\cal Q}_{\Sigma}\;=\;4\pi G \gaverage{\varrho} \;\;\;;\;\;\;
\gaverage{\CR} \;=\;12\pi G \gaverage{\varrho}  - 
6H_{\Sigma}^2  \;\;;\;\;H_{\Sigma}=\frac{\cal C}{a_{\Sigma}}\;\;,\;\;{\cal C} = const.\;\;,
\end{equation}
with the {\it global kinematical backreaction} $Q_{\Sigma}$, the globally averaged 
3--Ricci curvature $\gaverage{\CR}$, and the total restmass of the compact universe model 
$M_{\Sigma}$. The second condition above has been obtained by eliminating the 
backreaction term in  (\ref{averagehamilton}) using the first condition. 
Eliminating instead the density source we can determine the constant $\cal C$ by evaluating
(\ref{averagehamilton}) at the initial time: 
${\cal C}^2 := \frac{1}{2} {\cal Q}_{\Sigma}(t_i) - \frac{1}{6}\gaverage{\CR}(t_i)$.
The cosmic equation of state for a stationary cosmos can be obtained
by inserting (\ref{stationary})  into (\ref{equationofstate}). We find\footnote{It is
interesting to compare this condition with the investigation 
of backreaction in inhomogeneous cosmon fields by Christof Wetterich \cite{wetterich}, 
in particular with the  `cosmon equation of state'.}:
\begin{equation}
\label{cosmicstate1}
{\bf p}^{\Sigma}_{\bf\rm eff}\;=\; 
-\frac{1}{3}\;\boldsymbol{\varrho}^{\Sigma}_{\bf\rm eff}\;\;.
\end{equation}
Now, taking the time--derivative of the two conditions in (\ref{stationary})
and employing restmass conservation on $\Sigma$,
$\partial_t \gaverage{\varrho} + 3 H_{\Sigma}\gaverage{\varrho}=0$, 
we obtain evolution equations for the global kinematical backreaction and the
globally averaged Ricci curvature. These evolution equations are solved by:
\begin{equation}
\label{stationaryS}
{\cal Q}_{\Sigma} \;=\; \frac{{\cal Q}_{\Sigma}(t_i)}{a_{\Sigma}^{3}}\;\;\;;\;\;\;
\gaverage{\CR} \;=\;  \frac{\gaverage{\CR}(t_i)-3{\cal Q}_{\Sigma}(t_i)}{a_{\Sigma}^{2}} 
\;+\;\frac{3{\cal Q}_{\Sigma}(t_i)}{a_{\Sigma}^{3}}  \;\;,
\end{equation}
which indeed points to
a strong coupling between kinematical backreaction and averaged scalar curvature:
the rate of decay of  ${Q}_{\Sigma}$ is in proportion to $\gaverage{\varrho}$ 
and can therefore be of the same order as $4\pi G\gaverage{\varrho}$ today.
The total kinematical backreaction ${Q}_{\Sigma}V_{\Sigma} = 4\pi G M_{\Sigma}$ 
is a conserved quantity. Actually, the solution (\ref{stationaryS}) provides the first
example of an exact solution of the effective Einstein equations with a non--trivial
coupling of averaged scalar curvature to kinematical backreaction;
(\ref{stationaryS}) solves the integrability condition (\ref{integrability}).
  
As in Friedmannian cosmology the fate of the 
effectively stationary inhomogeneous cosmos is determined by initial conditions,
for the stationary effective scale factor 
$a_{\Sigma} = a_S + {\cal C}(t-t_i)$ can be restricted to a `Big Bang model'
by setting $a_S = 0$, or it can emerge \cite{ellismaartens} from the effectively static cosmos 
$a_S = const.$, which is a subcase of the stationary one by setting $H_{\Sigma}=0\;;\; 
{\cal C}=0\;\Leftrightarrow\; \gaverage{\CR}(t_i)\;=\;3{\cal Q}_{\Sigma}(t_i)$.
The globally static inhomogeneous cosmos is characterized by the cosmic equation of state:
\begin{equation}
\label{cosmicstate2}
\gaverage{\CR}\;=\;3{\cal Q}_{\Sigma}\;=\;const.\;\;\Rightarrow\;\;
{\bf p}^{\Sigma}_{\bf\rm eff}\;=\; \boldsymbol{\varrho}^{\Sigma}_{\bf\rm eff}
\;=\;0\;\;.
\end{equation}

\section{Concluding remarks}

We  argued that an explanation of  Dark Energy  by kinematical
backreaction effects and, hence, requiring the condition (\ref{accelerationcondition})
to hold in our Hubble volume, implies a cosmos that is dominated by strong
expansion fluctuations. 
We may speak of a {\it far--from--equilibrium} cosmic state in contrast to a
perturbed Friedmannian state. We further argued that 
a system featuring strong fluctuations
would conceivably, with some probability, create regional Hubble volumes in the vicinity
of the stationary state (\ref{stationaryS}). We have noted that, even if the 
Universe started in the vicinity of such a state, then it is important for the survival of
the condition (\ref{accelerationcondition}) that the evolution
implies strong coupling between kinematical backreaction and averaged scalar curvature.
We have shown that the globally stationary state indeed conserves 
strong fluctuations: the key--property is that ${\cal Q}_{\Sigma} \propto a_{\Sigma}^{-3}
\propto \gaverage{\varrho}$ in large contrast to the case of a perturbed
Friedmannian cosmos that would likely
evolve into small fluctuations on the Hubble scale.

The picture that emerges entails a regionally fluctuating cosmos, where the region is
as large as our Hubble volume. In fact this would imply a paradigmatic change of
cosmology,
since the properties seen on the Hubble scale are not extrapolated to the global
scale as in Friedmannian cosmology. This remark is more evident, if a globally
static model is envisaged: a globally static, but inhomogeneous cosmos 
is conceivable without employing a compensating cosmological constant in contrast to the
classical Einstein cosmos. This mathematical
possibility (put into perspective in \cite{buchert:jgrg}) 
would attain the status of a viable physical model, if 
the ``classical'' explanation of the Dark Energy problem in terms of
kinematical backreaction effects were true.
In such a cosmos the averaged scalar curvature  is, for a non--empty Universe, 
positive; the global kinematical backreaction term takes the role of a positive cosmological 
constant. Regionally, such a cosmos features exponential gravitational 
instabilities, i.e. it has strong matter and curvature fluctuations.

In a forthcoming work \cite{buchert:static} further implications are investigated.
In particular, it is argued that, e.g. a globally static dust cosmos, provided it originates in this state, 
could be stabilized by 
backreaction and averaged curvature in contrast to the global stability properties of the 
classical Einstein cosmos \cite{barrowetal:static}.
It is further clarified that a fluctuation--dominated cosmos, expressed in a thermodynamic 
language, is in a {\it far--from--equilibrium state}
compared with the Friedmannian {\it ``equilibrium state''} by employing an 
entropy measure proposed in \cite{hosoya:infoentropy},
which vanishes for Friedmannian cosmologies (``zero structure'') and is positive and time--dependent
for the globally stationary, but inhomogeneous cosmos.

\medskip

{\footnotesize\it This work was supported by the Sonderforschungsbereich SFB 375 
`Astroparticle physics' by the German science foundation DFG.
Special thanks go to St\'ephane Colombi and Sabino Matarrese for stimulating discussions.}

\end{document}